\documentclass[aps,prl,reprint,amsmath,amssymb]{revtex4-1}

\usepackage[english]{babel}
\usepackage[T1]{fontenc}
\usepackage[utf8]{inputenc}
\usepackage{xspace}
\usepackage{graphicx}
\usepackage{bm}
\usepackage{epstopdf} 
\usepackage{hyperref}
\hypersetup{
    colorlinks=true,
    citecolor=blue,
    filecolor=red,
    linkcolor=blue,
    urlcolor=blue,
    pdfborderstyle={/S/U},
}

\begin{document}


\title{Statistical Dynamics of Spatial-order Formation by Communicating Cells}


\author{Eduardo P. Olimpio}
\affiliation{Kavli Institute of Nanoscience}
\affiliation{Departments of Bionanoscience and Physics, Delft University of Technology, 2629HZ Delft, The Netherlands}
\author{Yiteng Dang}
\affiliation{Kavli Institute of Nanoscience}
\affiliation{Departments of Bionanoscience and Physics, Delft University of Technology, 2629HZ Delft, The Netherlands}
\author{Hyun Youk}
\email[]{h.youk@tudelft.nl}
\affiliation{Kavli Institute of Nanoscience}
\affiliation{Departments of Bionanoscience and Physics, Delft University of Technology, 2629HZ Delft, The Netherlands}


\begin{abstract}
Communicating cells can coordinate their gene expressions to form spatial patterns. \textquoteleft Secrete-and-sense cells\textquoteright \ secrete and sense the same molecule to do so and are ubiquitous. Here we address why and how these cells, from disordered beginnings, can form spatial order through a statistical mechanics-type framework for cellular communication. Classifying cellular lattices by \textquoteleft macrostate\textquoteright \ variables - \textquoteleft spatial order parameter\textquoteright \ and average gene-expression level - reveals a conceptual picture: cellular lattices act as particles rolling down on \textquoteleft pseudo-energy landscapes\textquoteright \ shaped by a \textquoteleft Hamiltonian\textquoteright \ for cellular communication. Particles rolling down represent cells' spatial order increasing. Particles trapped on the landscapes represent metastable spatial configurations. The gradient of the Hamiltonian and a \textquoteleft trapping probability\textquoteright \ determine the particle's equation of motion. This framework is extendable to more complex forms of cellular communication.
\end{abstract}


\maketitle

\begin{figure}[ht]
\includegraphics[width = 0.9\columnwidth]{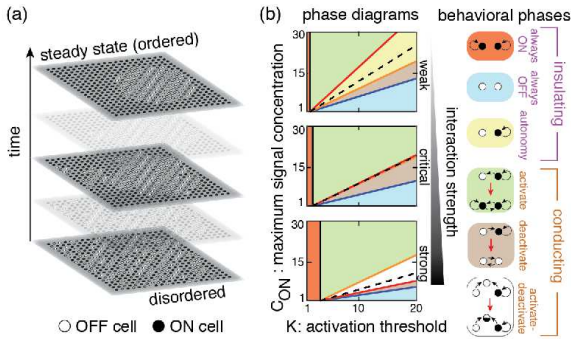}%
\caption{\label{fig:fig1} Behavioral phases of secrete-and-sense cells. (a) Snapshots of disorder-to-order transition dynamics. (b) (Left column): Phase diagrams for weak interaction (top), critical interaction (middle), and strong interaction (bottom). (Right column): Distinct behavioral phases in different colors.}
\end{figure}

Cells can communicate by secreting signaling molecules and this often underlies their collective behaviors. A striking example is initially uncoordinated cells, through cell-cell communication, coordinating their gene expressions to generate spatial patterns or structures \cite{Gregor:2010hh, Sawai:2005bq, Danino:2010jj, Liu:2011ei}. Many cells partly or completely control such \textquoteleft disorder-to-order\textquoteright \ dynamics by simultaneously secreting and sensing the same signaling molecule \cite{Doganer:2015ig, Youk:2014km}. These \textquoteleft secrete-and-sense cells\textquoteright \ appear across diverse organisms and include quorum-sensing social amoeba \emph{Dictyostelium} that form fruiting bodies \cite{Gregor:2010hh, Sawai:2005bq, Sgro:2015dm} and autocrine-signaling T-cells \cite{Antebi:2013ja, Sporn:1980ik}. Based on mounting evidence from studies of different organisms \cite{Gregor:2010hh, Sawai:2005bq, Danino:2010jj, Liu:2011ei, Youk:2014km, Sgro:2015dm, Antebi:2013ja, Mehta:2009fb, Kamino:2017bp, Hart:2014fy, DeMonte:2007cl, Umeda:2004es, You:2004de, Pai:2012hf, Coppey:2007fp, Shvartsman:2001bw}, researchers now suspect that secrete-and-sense cells, many of which are governed by the same core genetic-circuit architecture \cite{suppInfo, Doganer:2015ig}, are highly suited for spatially coordinating gene expressions. But if true, exactly why this is so, whether there are common design principles shared by different organisms, what the dynamics underlying their disorder-to-order transition is, and how to even quantify their spatial order, remain open questions. In this letter, we address these questions in the context of initially disordered fields of secrete-and-sense cells that self-organise into spatially ordered fields without any pre-existing morphogens. Specifically, we develop a theoretical framework that takes a simple and ubiquitous class of secrete-and-sense cells, sensibly defines and quantifies the notion of the cells' spatial order, and then elucidates how the spatial order evolves over time using analytical methods. We focus here on analytically describing how spatial correlations among cells' gene-expression levels dynamically emerge rather than on describing the shapes, sizes, and formations of specific spatial patterns (e.g., stripes).\\
\indent Our main idea is that describing hundreds of secrete-and-sense cells forming a particular spatial configuration is infeasible without exhaustive numerical simulations but that it is possible to analytically describe how an ensemble of \textquoteleft similar\textquoteright \ spatial configurations evolves over time without knowing the state of each cell. We will define a \textquoteleft spatial order parameter\textquoteright \ - a number between zero (complete disorder) and one (complete order). Inspired by statistical physics-approaches, we will group all lattices of cells that have the same spatial order parameter and average gene-expression level into an ensemble (\textquoteleft macrostate\textquoteright ). Surprisingly, we find that this macrostate moves like a particle that drifts-and-diffuses in an abstract space (\textquoteleft phase space\textquoteright ) whose coordinates denote the cells' spatial order and average gene-expression level. The particle drifts down a \textquoteleft pseudo-energy landscape\textquoteright \ defined by a \textquoteleft Hamiltonian\textquoteright \ for cell-cell communication that is akin to Hamiltonians of frustrated magnets. The gradient of the Hamiltonian and a \textquoteleft trapping probability\textquoteright \ that quantifies a \textquoteleft stickiness\textquoteright \ of the pseudo-energy landscape lead to an equation of motion in the phase space. We thus provide an intuitive picture, based on measurable quantities, that is both practical and conceptual for elucidating how cells spatially coordinate their gene expressions.\\
\indent We used a cellular automaton \cite{Ermentrout:1993ky} to simulate secrete-and-sense cells. We will compare its results with our theory. We considered a two-dimensional triangular lattice of $N$ spherical, immobile secrete-and-sense cells of radius $R$ and lattice spacing $a_o$. As a proof-of-principle, we considered \textquoteleft simple\textquoteright \ secrete-and-sense cells that (1) very slowly respond to their fast diffusing signal, and (2) whose gene expression level, which is determined by the extracellular concentration of the signal, and signal-secretion rate exhibit switch-like (digital) bistability \cite{suppInfo}. These two features were motivated by experimentally characterised secrete-and-sense cells. Examples include yeasts that secrete-and-sense a mating pheromone in a nearly digital manner (diffusion timescale $\sim$1 s; response timescale $\sim$30 minutes)\cite{Youk:2014km, Rappaport:2012ji} and mouse hair follicles, which are secrete-and-sense organs that act as digital secrete-and-sense cells on a triangular lattice (diffusion timescale $\sim$12 hours; response timescale $\sim$1.5 days)\cite{Chen:2015hh}. Each cell's gene expression is either \textquoteleft ON\textquoteright \ (and its signal-secretion rate is \textquoteleft high\textquoteright )  or \textquoteleft OFF\textquoteright \ (and its signal-secretion rate is \textquoteleft low\textquoteright ). Each cell senses the steady-state signal-concentration $c$ on itself. If $c$ is larger (lower) than a threshold concentration $K$, then the cell is ON (OFF). When $N=1$, an ON-cell (OFF-cell) maintains a steady-state concentration $C_{ON}$ ($C_{OFF}$) on itself.  We set $C_{OFF}$=1. The cellular automaton computes the $c$ on every cell, synchronously updates each cell's state, and repeats this process until it reaches a steady-state in which no cell requires an update. By running the cellular automaton on randomly distributed ON- and OFF-cells, we observed that initially disordered lattices could indeed evolve into spatially ordered steady-state configurations such as islands of ON-cells (Fig.~\ref{fig:fig1}a)\cite{Maire:2015de}. 

To reveal how the disorder-to-order dynamics arises, we will analyse the cellular automaton in each of the cells' \textquoteleft behavioral phases\textquoteright \ that we described in a previous work (Fig.~\ref{fig:fig1}b) \cite{Maire:2015de, suppInfo}. To recap, the behavioral phases represent how one cell turns on/off another cell. They arise from self-communication (i.e., a cell captures its own signal) competing with neighbor-communication (i.e., a cell captures the other cells' signal). This competition is characterised by an \textquoteleft interaction strength\textquoteright , $f_{N}$($a_{o}$) $\equiv$ $\sum_{i,j}e^{R-r_{ij}}/r_{ij}$ (where $r_{ij}$ is the distance between cell-$i$ and cell-$j$ in units of $a_o$). It measures how much of the other cells' signal diffuses to each cell \cite{Maire:2015de, suppInfo}. Given an interaction strength, the $K$ and $C_{ON}$ determine the cells' behavioral phase. The values of $K$, $C_{ON}$, and $f_{N}(a_o)$ are held fixed and thus the cells' behavioral phase also remains unchanged over time. A behavioral phase is either an \textquoteleft insulating phase\textquoteright \ - in which no cell can turn on/off the other cells due to dominant self-communication - or a \textquoteleft conducting phase\textquoteright \ - in which cells can turn on/off the others due to dominant neighbor-communication (Fig.~\ref{fig:fig1}b). Regardless of the interaction strength, cells can operate in two conducting phases: (1) \textquoteleft activate phase\textquoteright \ - in which neighboring ON-cells can turn on an OFF-cell, and (2) \textquoteleft deactivate phase\textquoteright \ - in which neighboring OFF-cells can turn off an ON-cell. Additionally, when the interaction is weak (i.e., $f_{N}$($a_o$)  $<$ 1), cells can operate in \textquoteleft autonomy phase\textquoteright \ - an insulating phase in which a cell can stay ON/OFF regardless of the other cells' states. On the other hand, when the interaction is strong (i.e., $f_{N}$($a_o$)  $>$ 1), cells can operate in \textquoteleft activate-deactivate phase\textquoteright \ - a conducting phase in which the cells can both activate and deactivate others depending on their relative locations \cite{suppInfo}.

\begin{figure}[ht]
\includegraphics[width = 1.0\columnwidth]{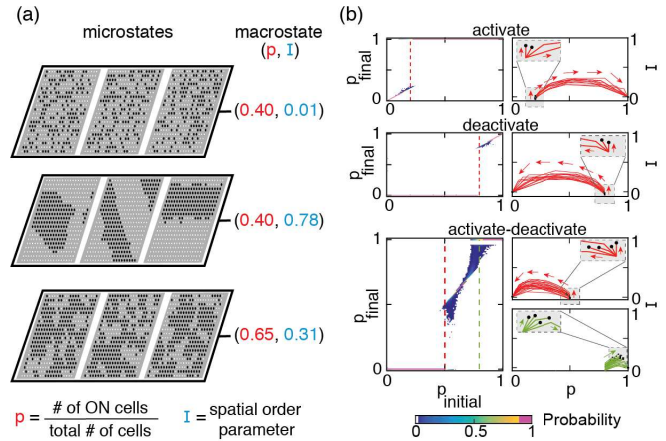}
\caption{\label{fig:fig2} Macrostates of secrete-and-sense cells. (a) Examples of microstates grouped into a macrostate. $p$ = average gene expression level (equivalent to fraction of cells that are ON), $I$ = spatial order parameter. ($p$, $I$) defines a macrostate. (b) (left column) Probability density maps of the particle's final value of $p$ (denoted  $p_{\text{final}}$) for every initial value of $p$ (denoted $p_{\text{initial}}$). (right column) Trajectories (red and green curves) in $p-I$ space in the activate phase (top), deactivate phase (middle), and activate-deactivate phase (bottom). Grey insets show zoomed-in views of trajectories. Black dots denote trajectories' endpoints.} 
\end{figure}
\begin{figure}[ht]
\includegraphics[width = 1.0\columnwidth]{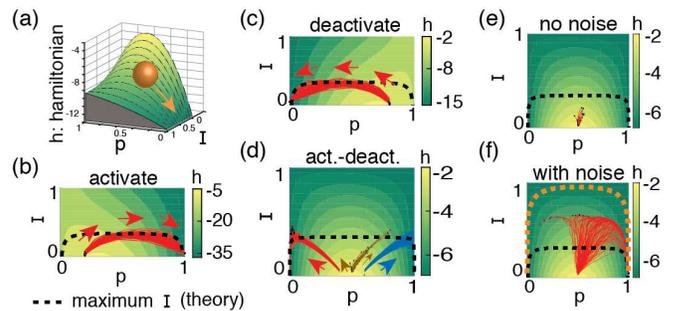}
\caption{\label{fig:fig3} Cellular lattices act as particles that roll down on pseudo-energy landscapes. (a) Pseudo-energy landscape defined by Hamiltonian $h(p, I)$ for cellular communication. Pseudo-energy landscape for (b) activate phase, (c) deactivate phase, and (d-f) activate-deactivate phase. (b-f) Trajectories of the same color start from the same position in each landscape. Black curves show maximally allowed $I$ - $I_{max}(p)$ \cite{suppInfo}. Trajectories without noise (e) and with noise (f) on the same landscape. Orange curve in (f) is the $I_{max}(p)$ when there is moderate noise \cite{suppInfo}.}
\end{figure}

	We now present our theory's central ingredient. Let us define two \textquoteleft macrostate\textquoteright \ variables: (1) the fraction $p$ of cells that are ON (equivalent to the average gene expression level) and (2) a \textquoteleft spatial order parameter\textquoteright \ $I$ defined as
\begin{equation}
I \equiv \frac{N}{\sum_{i, j} f\left(r_{ij}\right)} \frac{\sum_{i, j}  f\left(r_{ij}\right) \left(X_i - \langle X\rangle \right) \left(X_j - \langle X\rangle\right)}{\sum_i \left(X_i - \langle X\rangle\right)^2}
\label{eq_moranI}
\end{equation}
where $f(r_{ij})\equiv e^{R-r_{ij}}/r_{ij}$ is the cell-pair-$ij$'s interaction term, and $X_i$ is +1 (-1) for an ON (OFF)-cell.  By definition, $0 \leq \lvert I\rvert \leq 1$ and $ 0 \leq p \leq 1$. Roughly speaking, the $I$ measures the average correlation between any two cells by weighing each cell pair by its interaction strength \cite{Maire:2015de, MORAN:1950tj, suppInfo}. As $\lvert I\rvert$ approaches 0, the lattice becomes more disordered. As $\lvert I\rvert$ approaches 1, the lattice becomes more ordered. When $I=0$, ON- and OFF-cells are randomly distributed, yielding maximally disordered lattices. Our central idea is to group cellular lattices that have the same ($p$, $I$) into an ensemble (i.e., macrostate) (examples in Fig.~\ref{fig:fig2}a and \cite{suppInfo}). We then view a macrostate as a particle that moves in an abstract space (i.e., \textquoteleft phase space\textquoteright ) whose position at time $t$ is ($p(t)$, $I(t)$). Each point ($p$, $I$) in this phase space represents an ensemble of thousands of spatial configurations (\textquoteleft microstates\textquoteright ) \cite{suppInfo}. By randomly choosing thousands of microstates that all belong to the same disordered macrostate ($p = p_{\text{initial}}$, $I \approx 0$) and then running the cellular automaton on each one, we observed how the lattices evolved out of disorder (Fig.~\ref{fig:fig2}b). Specifically, we obtained a distribution of their trajectories (thus also a distribution of their final positions ($p_{\text{final}}, I_{\text{final}}$)) for each value of $p_{\text{initial}}$ in each behavioral phase (Fig.~\ref{fig:fig2}b) \cite{suppInfo}. The fact that we obtained distributions of trajectories for each $p_{\text{initial}}$, instead of a single trajectory, highlights that the particle moves stochastically. This arises from the automaton operating on individual cell's state $X_i$ - a microstate variable that we ignore - rather than on the macrostate variables, $p$ and $I$.  We observed that in every trajectory, the $I$ initially increased before plateauing while the $p$ either increased or decreased over time (Fig.~\ref{fig:fig2}b). Then, one of two events occurred: either (A) the particle stopped before its $p$ reached an extreme value (i.e., zero or one) (e.g., green trajectories in the activate-deactivate phase in Fig.~\ref{fig:fig2}b); or (B) the particle reached an extreme value of $p$ and as it did so, its $I$ abruptly dropped to zero (e.g., red trajectories in Fig.~\ref{fig:fig2}b). To explain observation (B), we first rewrite Eq. (\ref{eq_moranI}) as
\begin{equation}
I = \frac{\langle\sum_{i,j} f(r_{ij}) X_i X_j\rangle - (2p-1)^2f_N(a_o)}{4p(1-p)f_N(a_o)}.
\label{eq_moranI_in_p}
\end{equation}
From a mean-field approximation, we find that this quantity is bounded above by a function $I_{max}(p)$ that is less than or equal to 1 (dashed black curves in Fig.~\ref{fig:fig3}) \cite{suppInfo}. Accordingly, as $p$ nears zero or one, we find that the maximally allowed $I$ sharply decreases to zero (because $I_{max}(p)$ does as well - see Fig.~\ref{fig:fig3}) \cite{suppInfo}. This explains why the particle's $I$ abruptly drops to zero as its $p$ reaches zero or one (Fig.~\ref{fig:fig2}b). Yet, here the cells are becoming more correlated since every cell is either turning on ($p$=1) or off ($p$=0). Thus, we must interpret the $I$ carefully near the extreme values of $p$. To fully explain the particle trajectories along with observations (A) and (B), we next sought an equation of motion for the particles. 

 We conjectured that if a cellular lattice indeed moves like a particle, then there may be a landscape on which the particles roll down. To explore this idea, we define a \textquoteleft multicellular Hamiltonian\textquoteright , $h \equiv -\sum_i X_i \left(Y_i - K \right)/N$, where $Y_i$ is the signal concentration on cell-$i$. In fact, we can rewrite $h$ entirely in terms of the macrostate variables, $p$ and $I$ \cite{suppInfo}. Plotting $h(p, I)$ yields a \textquoteleft pseudo-energy landscape\textquoteright \ (Fig.~\ref{fig:fig3}a). Its shape depends on the cells' behavioral phase (Figs.~\ref{fig:fig3}b--\ref{fig:fig3}d). Importantly, by plotting the trajectories on top of their respective landscapes, we observed that every particle's pseudo-energy monotonically decreased over time until the particle stopped - a fact that we could also rigorously prove \cite{suppInfo}. Crucially, in every behavioral phase, the pseudo-energy landscape slopes downwards towards increasing $I$ (Figs.~\ref{fig:fig3}b--\ref{fig:fig3}d). The particles cannot roll all the way down due to the existence of the $I_{max}(p)$ (Figs.~\ref{fig:fig3}b--\ref{fig:fig3}d). To see at the microstate-level why the cells become more spatially correlated over time, we rewrite the $h$ as
\begin{equation}
h = -\alpha\sum_{i, j}f(r_{ij}) X_i X_j - B\sum_i X_i  - N\alpha ,
\label{eq_hamiltonian_sums}
\end{equation}
where $\alpha \equiv (C_{ON}-1)/(2N)$, and $B$ is a \textquoteleft signal field\textquoteright \ identified as $\alpha(1+f_N(a_o))-K/N$.  Eq.~\ref{eq_hamiltonian_sums} is strikingly similar to the Hamiltonians of the Hopfield network \cite{Hopfield:1982vm} and magnetic spins with long-range interactions \cite{Kirkpatrick:1975vi}. As in physical systems, the signal field is a knob that an experimentalist (and cells) can tune to sculpt the pseudo-energy landscape \cite{suppInfo}. It competes with the cell-cell interaction term in Eq. \ref{eq_hamiltonian_sums} (with coupling constant $\alpha f(r_{ij})$). From the phase diagrams, we can deduce that $B > 0$ in the activate phase, that $B < 0$ in the deactivate phase, and that $B$ can be positive, negative, and zero in the activate-deactivate phase (depending on $K$ and $C_{ON}$) \cite{suppInfo}. Intuitively, increasing $\langle\sum_{i,j} f(r_{ij}) X_i X_j\rangle$, and thus the $I$ (by Eq. (\ref{eq_moranI_in_p})), corresponds to more clusters of ON-cells and OFF-cells forming, which would in turn decrease the Hamiltonian since averaging the first term in Eq.~\ref{eq_hamiltonian_sums} yields $-\alpha\langle\sum_{i,j} f(r_{ij}) X_i X_j\rangle$. 

\begin{figure}[ht]
\includegraphics[width = 0.9\columnwidth]{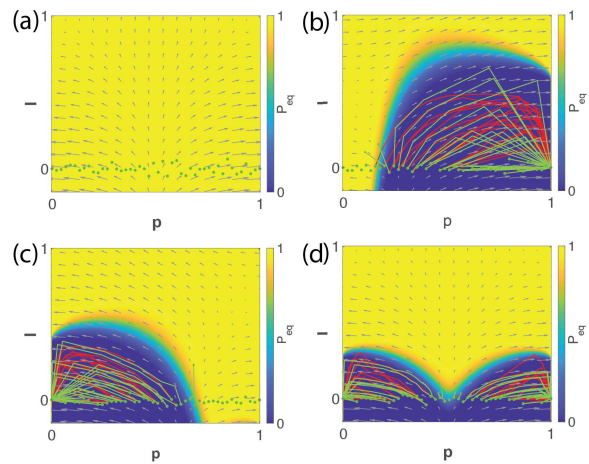}
\caption{\label{fig:fig4} Gradient field of the multicellular Hamiltonian, $-\vec{\nabla} h(p, I)$, and equilibrium probability $P_{eq}$ determine particle trajectories and the equation of motion. (a-d) An arrow at ($p$, $I$) is the gradient of the multicellular Hamiltonian at that point. Heat maps show magnitude of the trapping probability $P_{eq}$ at each location. Red trajectories are exact particle trajectories from the cellular automaton. Green trajectories are particle trajectories produced by the equation of motion (Eq.~\ref{eq_motion}). Autonomy phase (a), activate phase (b), deactivate phase (c), and activate-deactivate phase (d).}
\end{figure}

	To deduce how exactly the shape of the pseudo-energy landscape determines the particle's motion and obtain the equation of motion, we compared the gradient field of the multicellular Hamiltonian, $\vec{V} (p, I)=-\vec{\nabla} h$ (arrows in Figs.~\ref{fig:fig4}a--\ref{fig:fig4}d) with the particle trajectories produced by the cellular automaton (red curves in Figs.~\ref{fig:fig4}a--\ref{fig:fig4}d). We discovered that the particles closely follow the streamlines dictated by the gradient field. From this and our observation that the particles follow stochastic trajectories, we conjectured that the particles follow a Langevin-type dynamics in which the particle drifts (rolls) down the pseudo-energy landscape due to the gradient field $\vec{V}$ and diffuses due to a noise term:
\begin{equation}
(\Delta p(t), \Delta I(t)) = -\vec{\nabla} h(p(t), I(t)) \, \delta +  (\eta_p(t), \eta_I(t)),
\label{eq_motion}
\end{equation}
where $ \Delta p(t)$ and $\Delta I(t)$ are changes in $p$ and $I$ respectively at time step $t$, $\delta$ is a constant factor that scales the gradient to account for the discreteness of time in the cellular automaton, and $\eta_p(t) \sim \mathcal{N}(0,\sigma_p)$ and $\eta_I(t) \sim \mathcal{N}(0,\sigma_I)$ are white noise terms representing our ignorance of the microstates. We determined $\delta, \sigma_p$ and $\sigma_I$ by calculating the mean and the variance of $\Delta p$, which in turn are set by the distribution of signal concentrations that a cell senses for a given $(p,I)$ \cite{suppInfo}. We found that the particle trajectories obtained from Monte Carlo simulations of Eq.~\ref{eq_motion} (green curves in Fig.~\ref{fig:fig4}) closely match the exact particle trajectories  dictated by the cellular automaton (red curves in Fig.~\ref{fig:fig4}) for a wide range of parameters despite some deficiencies.

 Unlike physical energy landscapes, the pseudo-energy landscape can trap particles at its sloped regions, particularly but not exclusively in the activate-deactivate phase and the autonomy phase (e.g., brown trajectories in Fig.~\ref{fig:fig3}d and the red trajectories in Fig.~\ref{fig:fig3}e). These trajectories all terminate at intermediate values of $p$, before reaching the absolute minima of $h$. The trapped particles represent microstates that are steady states. By computing the fraction $P_{eq}(p,I)$ of steady microstates for each macrostate ($p$,$I$), we could predict and understand where the particles get trapped \cite{suppInfo}. For a moving particle at location ($p$,$I$), the \textquoteleft trapping probability\textquoteright \ $P_{eq}(p,I)$ represents a probability that the particle gets trapped at that location - it represents a \textquoteleft stickiness\textquoteright \ of the landscape. Plotting the $P_{eq}$ as a heat map on top of the gradient field (Fig.~\ref{fig:fig4}) yields a complete picture of particle's direction of motion (due to $\vec{V}$) and where it gets stuck (due to $P_{eq}$). Given an initial position in $(p,I)$ space, we can follow the vector field until the particle stops, which either occurs in a region of high $P_{eq}$ (yellow in Fig. \ref{fig:fig4}) or at the extreme values of $p$. At the microstate-level, the trappings arise due to a geometric restriction that is similar in flavor to, but not the same as, the geometric frustrations in magnetic spins \cite{Tchernyshyov:2011wj}. For example, in certain regions of the activate-deactivate phase, if we consider a lattice consisting of just five ON-cells in a cluster, we can show that for an OFF-cell to turn on, at least three of its nearest neighbors must be on whereas for an ON-cell to turn off, at least five of its nearest neighbors must be on \cite{suppInfo}. Since the lattice is triangular, neither of the two conditions can be met even though both would decrease the pseudo-energy. Hence the particle that represents this lattice would be trapped.

Finally, as a proof-or-principle for showing how to include stochastic gene expressions \cite{Sagues:2007fe, GarciaOjalvo:2011hm, Tkacik:2011jr, Sanchez:2013gg, Xu:2016kd, Friedman:2006}, we added stochastic sensing to the cellular automaton. We allowed the $K$ to fluctuate around a mean $K_o$ with a standard deviation $\delta K$, from cell to cell and from time to time. We found that noise could liberate trapped particles \cite{suppInfo}. In this sense, the trapped particles represent metastable spatial configurations. In particular, we found that a moderate noise, which occurs when $\delta K \sim$ min$(\lvert \langle Y_{ON}\rangle-K_o\rvert, \lvert \langle Y_{OFF}\rangle -K_o\rvert)$ (where $\langle Y_{ON(OFF)}\rangle$ is the mean signal-concentration on an ON (OFF)-cell), could liberate trapped particles and push them further down the landscape, beyond the previously allowed region, until they became trapped in regions of higher spatial order (compare Fig.~\ref{fig:fig3}f with Fig.~\ref{fig:fig3}e). These particles also cannot roll to the bottom because, just as when the sensing was deterministic, we found that an $I_{max}(p)$ still exists when a moderate noise is present (orange curve in Fig.~\ref{fig:fig3}f) \cite{suppInfo}. Intriguingly, we observed that some of these newly trapped particles' $p$, $I$, and $h$ changed very slowly, allowing them to remain trapped at an intermediate $p$ over hundreds but not thousands of time steps \cite{suppInfo}. We expect a follow-up study to examine if this is related to glass-type dynamics.

 Here we uncovered a visual landscape for cellular communication and showed that it underlies why and how simple secrete-and-sense cells' gene expressions become more spatially correlated over time. In the process, we revealed connections between secrete-and-sense cells, drifting-and-diffusing particles, and magnetic spins by defining quantities whose names originate from statistical mechanics but whose meanings are adapted to describe gene expressions of communicating cells. The theory does not yet account for more complex forms of secrete-and-sense cells but may be extended to do so. Towards this end, we show in the Supplementary Material \cite{suppInfo} how to extend our framework to lattices with multiple cell-types and signal-types, including paracrine-signaling \cite{Doganer:2015ig, infoParacrine}. We hope that our work, along with complementary approaches for studying spatial patterns \cite{Surkova:2009cg, Sokolowski:2012bxa, Tkacik:2008ht, Hillenbrand:2016ie, Erdmann:2009hwa, Cotterell:2010hk, Cotterell:2015cd, Fancher:2017ba, Thalmeier:2016ig}, will inform on-going efforts to establish generic frameworks for multicellular gene regulations. 

We thank Pieter Rein ten Wolde, Arjun Raj, Louis Reese, Yaroslav Blanter, and members of Youk group for discussions. HY was supported by the European Research Council and the Dutch Organization for Scientific Research.

\bibliography{Olimpio_Dang_Youk}

\begin{thebibliography}{43}
\bibitem [{inf()}]{infoParacrine}%
  \BibitemOpen
  \href@noop {} {}\bibinfo {note} {In paracrine signaling, one cell secretes a
  signal without sensing it while another cell senses without secreting that
  signal.}\BibitemShut {Stop}%
\end{thebibliography}
\end{document}